# Interpretable Machine Learning Model for Early Prediction of Mortality in Elderly Patients with Multiple Organ Dysfunction Syndrome (MODS): a Multicenter Retrospective Study and Cross Validation

Xiaoli Liu[1,4*], Pan Hu[2,3*], Zhi Mao[2], Po-Chih Kuo[4,5], Peiyao Li[6,7], Chao Liu[8], Jie Hu[2,9], Deyu Li[1], Desen Cao[7], Roger G. Mark[4,10], Leo Anthony Celi[4,10], Zhengbo Zhang[7,11,1*], Feihu Zhou[2,12*]

liuxiaoli@buaa.edu.cn, hupan0215@163.com, maozhi@126.com, bfantasykuo@gmail.com, lipy19@mails.tsinghua.edu.cn, chaoliu301@foxmail.com, catherinehujie@163.com, deyuli@buaa.edu.cn, caodesen2012@126.com, rgmark@mit.edu, leoanthonyceli@yahoo.com, zhengbozhang@126.com, feihuzhou301@126.com

## Summary

**Background:** Elderly ICU patients with multiple organ dysfunction syndrome (MODS) have a high risk of death therefore early detection is critical for an improved prognosis. The performance of current scoring systems used to assess the severity of MODS lack sufficient sensitivity and specificity to accurately guide clinicians' actions. This study aims to develop an interpretable and generalizable model with superior performance for early mortality prediction in elderly patients with MODS.

**Methods:** The MIMIC-III, eICU-CRD and PLAGH-S databases (with 57,786, 200,859 and 1,462 unique ICU admissions, respectively) were employed for model generation and evaluation. We used the machine learning model XGBoost (eXtreme Gradient Boosting) with the SHapley Additive exPlanations method to conduct early and interpretable predictions of patients' hospital outcome. Three types of data source combinations (from single center, multicenter and the fusion of both) and five typical evaluation indexes (AUC, sensitivity, specificity, F1 and accuracy) were adopted to develop a generalizable model.

**Findings:** The interpretable model, with optimal performance developed by using MIMIC-III and eICU-CRD datasets, was separately validated in MIMIC-III, eICU-CRD and PLAGH-S datasets (no overlapping with training set). The performances of the model in predicting hospital mortality as validated by the three datasets were: AUC of 0·858, sensitivity of 0·834 and specificity of 0·705; AUC of 0·849, sensitivity of 0·763 and specificity of 0·784; and AUC of 0·838, sensitivity of 0·882 and specificity of 0·691, respectively. Comparisons of AUC (95% CI) between this model and baseline models with MIMIC-III dataset validation showed superior performances of this model (0·858 [0·841 - 0·875] vs. 0·854 [0·838 - 0·871], 0·834 [0·817 - 0·853], 0·824 [0·804 - 0·844], 0·788 [0·766 - 0·810] and 0·741 [0·716 - 0·765] compared with LR, NN, SVM, RF and NB, respectively); In addition, comparisons in AUC between this model and commonly used clinical scores showed significantly better performance of this model (0·858 [0·841 - 0·875] vs. 0·752 [0·729 - 0·776], 0·73 [0·704 - 0·757], 0·694 [0·669 - 0·719], 0·686 [0·659 - 0·713] and 0·668 [0·639 - 0·697] compared with OASIS, APSIII, MODS, SAPS and SOFA, respectively).

**Interpretation:** The interpretable machine learning model developed in this study using fused datasets with large sample sizes was robust and generalizable. This model outperformed the baseline models and several clinical scores for early prediction of mortality in elderly ICU patients. The interpretative nature of this model provided clinicians with the ranking of mortality risk features and the rationale for assessing the patient's mortality risk probability.

# Introduction

Multiple organ dysfunction syndrome (MODS) is a continuous process with physiologic derangement in more than one organ,[1] and its leading culprits include infection, injury, hypoperfusion and hypermetabolism status, etc..[2] The high morbidity and mortality, as well as the substantial medical expenses in patients who are admitted to intensive care unit (ICU) and encounter MODS, have been a very challenging issue.[1,3] It should be noted that elderly patients (≥ 65 years old) with MODS have a significantly higher mortality risk compared with younger patients due to their fragile health status and potential comorbidities.[4,5] A prior study reported that the mortality risk in elderly patients suffering from multiple organ (over three) failure could be up to 50% ~ 100%.[6] Moreover, even if they survived, the need for long-term clinical care and organ function support treatment would be a heavy financial burden that is hardly bearable.[7,8] Therefore, early assessment of organ failure severity and mortality prediction in elderly patients with MODS are of vital importance in giving clinicians more time to respond by providing individualized clinical and nursing care.

Since 1980, extensive studies on clinical scores evaluating the risk of death based on patient's organ function or severity of illness have been carried out. These include the Acute Physiology and Chronic Health Evaluation-II (APACHE-II) score established by Knaus et al,[9] followed by the modified score of the APACHE III prognostic system developed in 1991,[10] Simplified Acute Physiology Score (SAPS II) including 17 variables (19 variables in the APACHE-III) in assessing the severity of organ failure proposed by Le Gall et al,[11] Multiple organ dysfunction score (MODS) by Marshall et al,[12] and the widely recognized Sequential Organ Failure Assessment (SOFA) score developed by Vincent et al.[13] However, a growing body of literature has demonstrated that the scores mentioned above failed to accurately assess and predict the risk of death[14] for the following reasons: 1) the factors and their assigned weights according to a panel of experts' experience could not fully reflect the characteristic of a larger population;[15,16] 2) the individual linear addition of each organ system could not represent the complex situation and the intrinsic correlations of organ systems;[16] and 3) they were not adequately calibrated in multicenter and large sample cohorts.[15]

Recently, the availability of electronic health records (EHR) data has allowed researchers to focus on developing machine learning algorithms for powerful analysis of complex and heterogeneous data and sophisticated modeling capacity.[17] The OASIS severity score, designed by Johnson et.al, was a novel illness scale using part of the variables from APACHE-III and particle swarm optimization algorithm to achieve an effective prediction of mortality and length of hospitalization.[18] The Super ICU Learner Algorithm (SICULA) was proposed by Romain et al to improve mortality prediction,[15] which adopted an ensemble machine learning method with a better predictive performance than those obtained from scoring systems. However, these scores or algorithms focus solely on adult ICU patients and lack sufficient samples for external validation of the model. Specifically, SICULA can only provide clinicians with risk probability but not the rationale for the assessment. Targeting this, Benjamin et. al developed an acuity assessing score, DeepSOFA, using deep learning methods.[19] With the same elements as a SOFA score, it was more optimal in evaluating disease severity by providing clinicians with a more accurate mortality prediction than its predecessor. However, it only included limited information without exploring other potentially meaningful factors for diagnosis. Meanwhile, the model was developed based on a local database and validated in a public single-center database from the same country, without further analysis on how to get a robust and universal prediction model. A Meyer et. al also employed deep learning methods to develop a real-time model of serious complications including mortality.[20] While it performed well, it is a black box to clinicians. Moreover, so far, few research studies focusing on elderly patients with MODS have been conducted.

In this paper, we aim to develop a prediction model to assist clinicians in the early diagnosis and treatment of this specific elderly population admitted to the ICU. With rigorous methodology, we propose an efficient way to acquire a robust and generalizable model, which is then validated in multicenter and cross-country datasets (developed: US, developing: China) with large sample sizes. The assessment of the degree of nervous system damage, as indicated in Glasgow coma score (GCS), was found to deserve more attention from clinicians. Likewise, two important risk factors, blood urea nitrogen (BUN) and shock index (SI), which are largely ignored in current commonly-used clinical scores, were identified to be highly relevant to patient mortality. The proposed model with interpretability can help clinicians better understand the decision-making process in the assessment of disease severity and take full advantage of any opportunities for early intervention.

## Methods

We performed a longitudinal, multicenter, retrospective study based on three high-volume databases, including the Medical Information Mart for Intensive Care Database v1·4 (MIMIC-III),[21,22] eICU Collaborative Research Database v1·2 (eICU-CRD) and PLAGH Surgical Intensive Care Database v1·1 (PLAGH-S).[23]

### Data Sources

MIMIC-III is a large, open-access, single-center dataset of 38,605 deidentified ICU patients who were admitted to the Beth Israel Deaconess Medical Center between 2002 and 2012. eICU-CRD is a multi-center, telehealth ICU and freely available dataset including over 200,000 de-identified ICU admissions covering 208 Critical Care Units in the United States from 2014 and 2015. PLAGH-S is a surgical ICU database under development from the General Hospital of the People's Liberation Army (PLAGH), a tertiary hospital integrating high-level education and research in China. We utilized data from 1,102 patients in 2017 and 2018, in which the identification information of patients was removed similar to the MIMIC-III database. These three databases (MIMIC-III, eICU-CRD and PLAGH-S) contain all medical records of patients during their ICU stay, typically including demographic information, vital signs, laboratory tests, medications, diagnoses, orders, notes, input and output information. It is worth mentioning that there might be variations in the method and frequency of vital sign data. For the eICU-CRD database, vital signs of periodic monitoring by machine were automatically interfaced, stored archived with a median interval of five minutes. In the MIMIC-III and PLAGH-S database, the vital sign values were recorded and confirmed by nurses. The recording frequency depends on the patient's disease severity. When the patient's condition was not stable, the nurse would record once every 5 to 15 minutes. Otherwise, it was usually recorded once per hour.

Contents in this study involving MIMIC-III and eICU-CRD databases was an analysis of a third-party anonymized publicly available database with pre-existing institutional review board (IRB) approval. Contents in this study using PLAGH-S database was approved by the ethics committee of the General Hospital of PLA (No·S2017-054-01).

### Study Population

All ICU patients 65 years old or older with multiple organ dysfunction syndrome, namely existing over two failure organs according to the sequential organ failure assessment score,[24] were included from three databases. We excluded patients with unknown outcomes, less than 24h of ICU admission stay, and secondary or multiple times entering hospital to avoid repeated inclusion. The patients without heart rate, respiratory rate, mean arterial pressure, glasgow coma score, temperature and oxygen saturation data in the first 24h ICU admission were also excluded.

**Data Extraction**

The following seven types of information representing the study cohort's baseline information were collected to develop the prediction model: 1) patient characteristics such as age, gender, ethnicity, diagnosis upon admission using International Classification of Diseases (9th revision); 2) clinical scores of the first day in ICU, reflecting patient's diseases severity, including APSIII acute physiology score III (APSIII)[10], OASIS and SOFA scores; 3) vital signs of the first day in ICU, including heart rate, respiratory rate, and mean arterial pressure, etc.; 4) laboratory results of the first day in ICU, including glucose, creatinine, white blood cell count, and bilirubin level, etc.; 5) the fluid input and urine output recorded in the first day in ICU; 6) the clinical treatments received during the first day in ICU, including mechanical ventilation, continuous renal replacement therapy and vasopressors usage; 7) in-hospital outcomes including mortality, duration of hospital and ICU stay. Table 1 presents an overview of the extracted information from each database. Detailed description is provided in the online supplementary files.

**Table 1: Information extracted from three databases involved in this study**

| Types | Contents |
|---|---|
| Demographic (18) | Age, gender, ethnicity, admission type, ICU admission unit, height, weight, chronic liver failure (CLF), acute liver failure (ALF), chronic obstructive pulmonary disease (COPD), acute respiratory distress syndrome (ARDS), coronary artery disease (CAD), chronic renal failure (CRF), acute renal failure (AKI), chronic heart failure (CHF), acute heart failure (AHF), stroke, and malignancy |
| Clinical scores (6) | APSIII, OASIS, SOFA, MODS, Systemic inflammatory response syndrome (SIRS) and GCS |
| Vital signs (8) | Heart rate (HR), respiratory rate (RR), mean arterial pressure (MAP), systolic blood pressure (SBP), diastolic blood pressure (DBP), central venous pressure (CVP), temperature (T) and oxygen saturation ($SpO_2$) |
| Laboratory tests (32) | Albumin, alkaline phosphatase, alanine transaminase (ALT), ast aspartate transaminase (AST), base excess (BE), prothrombin time (PT), partial thromboplastin time (PTT), bicarbonate, bilirubin, brain natriuretic peptide (BNP), BUN, chloride, creatinine, fibrinogen, glucose, hematocrit, hemoglobin, international normalized ratio (INR), lactate, lymphocytes, magnesium, neutrophils, partial pressure of oxygen ($PaO_2$), fraction of inspired oxygen ($FiO_2$), partial pressure of carbon dioxide in arterial blood ($PaCO_2$), $PaO_2/FiO_2$ ratio, ph, platelet, potassium, sodium, troponin, and white blood cell (WBC) |
| Output (1) | Urine output (UO) |
| Treatments (6) | Mechanical ventilation (vent), continuous renal replacement therapy (CRRT), vasopressors usage of dobutamine, dopamine, epinephrine, and norepinephrine |
| Outcomes (3) | Hospital outcome, days of hospital stay, and days of ICU stay |

**Feature Selection**

According to the initial data from three databases, we further dug more information to represent patients' diseases severity. Based on the seven types of records, the statistical features of maximum (max), minimum (min), average (avg), sum and mapping to 'Yes/No' were calculated by the extracted data expecting patient characteristics data. For the type of vital signs, the max, min and avg were entirely calculated of each sub-term; For the type of laboratory tests, only the max and min were obtained of each sub-term; For the type of treatment, vent, CRRT and vasopressors of dobutamine were represented by flag indicating whether a patient has received the corresponding treatment, and the max rate flow of other vasopressors were obtained for more specific information; For the type of output, urine output of the first 3, 6, 12 and 24 hours staying in ICU were acquired. Missing data was processed as follows: Missing values were imputed using the median value of each feature except for $FiO_2$ (with the imputation of 21%). Additionally, for the proportions of missing part greater than 30%, flag indication of whether a record existing was produced as new features. Finally, a total of 130 features were constructed. Additional information can be found in the supplementary materials. The proportions of missing raw data and details of the features are presented in supplementary eTable 4.

**Statistical Analysis**

All continuous variables including clinical scores were reported as medians with $25^{th}$ and $75^{th}$ interquartile ranges. The *t* test or Wilcoxon rank sum test was used when appropriate to compare between surviving and non-surviving elderly with MODS. Categorical variables were calculated to obtain the total number and percentage. P values (two sides) less than 0·05 was considered statistically significant.

**Model Development**

The eXtreme Gradient Boosting model (XGBoost) was employed for early assessment of patients' risk of death in hospital.[25] It's a machine learning algorithm with high computational speed and satisfactory prediction performance, due to its improvements of tree boosting in ensemble technique. Thousands of decision trees, weak learners generated with this method, are composed into strong learners using gradient boosting approach to iteratively train and optimize the parameters. In order to get a universal and robust prediction model, we explored the impact of different types of data sources on model performance. Considering the small samples size of the PLAGH-S cohort, it was suitable as a test set for evaluation. Therefore, we employed three models displayed in Figure 1. Model 1 was developed using 80% of a cohort randomly selected from the MIMIC-III database, which represented a single center database. Model 2 was developed from the eICU-CRD database using the same methods, which results in cohorts of patients from multiple centers. Model 3 was developed from the combined data from Models 1 and 2, which comprised a larger sample size and longer time span. For each model, the cross-validation process was not carried out in this paper owing to its large sample sets. We chose 'AUC' as the model's evaluation metric to reduce the bias of class imbalance. The important hyperparameters were set to default values, including the learning rate (learning_rate = 0·1), the maximum depth of each tree (max_depth = 3) and the numbers of modeling sequential trees (n_estimators = 100).

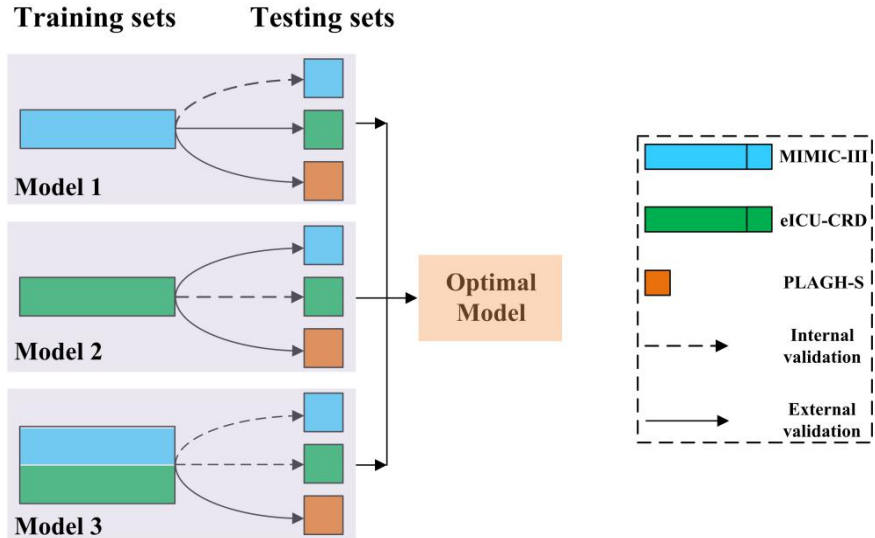

Figure 1: The process of developing the optimal prediction model

**Model Evaluations**

We obtained the three cohorts as testing sets from MIMIC-III (20% of data), eICU-CRD (20% of data) and PLAGH-S (all of data). Detailed analyses to assess the models' performances in different data sources were conducted, including internal and external validations, comparisons of baseline models and clinical scores (five machine learning methods: logistic regression (LR), support vector machine (SVM), neural network (NN), random forests (RF) and naive bayesian (NB); and five commonly used scores: SOFA, MODS, SAPS, OASIS and APSIII). The evaluation indexes included AUC, sensitivity, specificity, accuracy and F1 score.

**Model interpretation**

SHAP (SHapley Additive exPlanations) is a novel approach to explain various black box machine learning models, which had been validated in its interpretability performance and had been proven helpful for anesthesiologists to identify hypoxaemia during surgery.[26,27] The method defines the Shapley value as the only indicator to evaluate a feature's effect and adopts three properties with local accuracy, missingness and consistency to measure the feature's importance.[28] We leveraged it to provide the interpretation of our early prediction model with contributing risk factors leading to death in elderly patients with MODS.

The data extraction was accomplished with PostgreSQL Version 9·6. All calculations and analyses were performed utilizing Python Version 3·7·1 (the xgboost, sklearn and shap packages) and R Version 3·6·0 (the tableone package).

## Results
**Patient characteristics**

This study included 15,804 elderly patients (2,353 non-survivors) with MODS for analysis in MIMIC-III cohort. A total of 34,201 (3,966 non-survivors) and 439 (51 non-survivors) patients were respectively included in eICU-CRD and PLAGH-S cohorts, respectively. Figure 2 shows the inclusion criteria of three data sources' cohorts (see the supplementary files eFigure 1 to eFigure 3 for the detailed inclusion and exclusion criteria in each cohort. Table 2 summarizes the characteristics of three cohorts. The eTable 1 displays the comparison baseline information of survivors and non-survivors in the MIMIC-III study cohort. The non-survivors were significantly older in age and had higher severity scores and lower BMI upon ICU admission. Incidence of comorbidities such as COPD, AHF and CRF were not significantly different between the two groups. Duration of ICU stay was longer among non-survivors while duration of hospital stay was shorter among them. The patient characteristics of the eICU-CRD cohort was similar to MIMIC-III's (eTable 2), except the comorbidities were slightly different. Limited by the small size of the sample set, only differences in days of hospital/ ICU stay and severity scores of APSIII/ SOFA were consistent to the counterparts in PLAGH-S cohort (eTable 3).

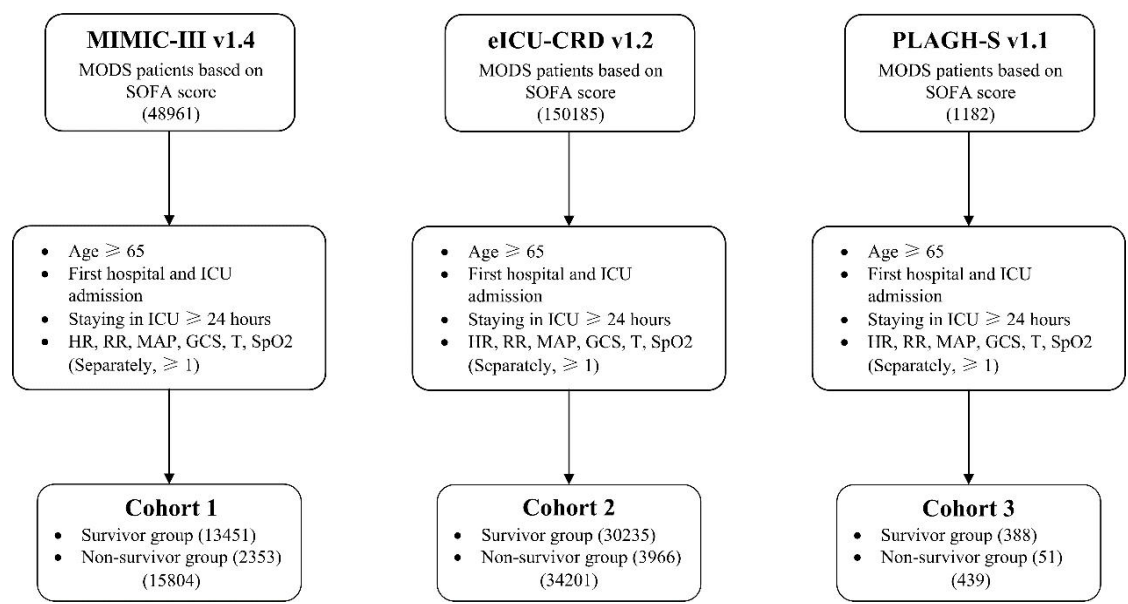

Figure 2: An overview of inclusion criteria with the multicenter study cohorts

Table 2: The comparison of the total study cohorts' baseline characteristic

|  | MIMIC-III (n = 15,804) | eICU-CRD (n = 34,201) | PLAGH-S (n = 439) |
|---|---|---|---|
| **Demographic** | | | |
| Age (yr), (median, IQR) | 77·00 [71·00, 83·00] | 76·00 [70·00, 83·00] | 73·00 [68·00, 79·50] |
| Male, n (%) | 8281 (52·4) | 17586 (51·4) | 248 (56·5) |
| Weight (kg), (median, IQR) | 77·20 [65·00, 90·70] | 77·30 [64·40, 92·00] | 65·00 [58·00, 72·00] |
| Height (cm), (median, IQR) | 167·64 [160·02, 175·26] | 167·60 [160·00, 177·00] | 165·00 [160·00, 172·00] |
| BMI (kg/m$^2$), (median, IQR) | 27·97 [24·37, 32·07] | 27·10 [23·30, 31·80] | 24·00 [22·00, 27·00] |
| ICU type (%) | | | |
| CCU | 2738 (17·3) | 7077 ( 20·7) | ·· |
| CSRU | 3509 (22·2) | 902 (2·6) | ·· |
| MICU | 5428 (34·3) | 21382 (62·5) | ·· |
| NICU | ·· | 2860 (8·4) | ·· |
| SICU | 2473 (15·6) | 1980 (5·8) | 439 (100·0) |
| TSICU | 1656 (10·5) | ·· | ·· |
| Ethnicity (%) | | | |
| ASIAN | 360 (2·3) | 386 (1·1) | 439 (100·0) |
| BLACK | 903 (5·7) | 2850 (8·3) | ·· |
| HISPANIC | 264 (1·7) | 1259 (3·7) | ·· |
| OTHER/UNKNOWN | 2504 (15·8) | 1541 (4·5) | ·· |
| WHITE | 11773 (74·5) | 28165 (82·4) | ·· |
| Admission type (%) | | | |
| ELECTIVE | 2332 (14·8) | 12974 (37·9) | ·· |
| FLOOR | ·· | 3400 (9·9) | 39 (8·9) |
| INTERMEDIATE CARE UNIT | ·· | 2347 (6·9) | ·· |
| OTHER/KNOWN | ·· | 9627 (28·1) | 259 (59·0) |
| URGENT | 466 (2·9) | 5853 (17·1) | ·· |
| EMERGENCY | 13006 (82·3) | ·· | 141 (32·1) |
| **Comorbidities** | | | |
| CLF | 436 (2·8) | 225 (0·7) | 8 (1·8) |
| ALF | 250 (1·6) | 58 (0·2) | 77 (17·5) |
| COPD | 3542 (22·5) | 4250 (12·4) | 37 (8·4) |
| ARDS | 2960 (18·8) | 7655 (22·4) | 66 (15·0) |
| CAD | 7129 (45·3) | 3069 (9·0) | 181 (41·2) |
| CRF | 2248 (14·3) | 3635 (10·6) | 31 (7·1) |
| AKI | 4165 (26·5) | 4514 (13·2) | 77 (17·5) |
| CHF | 5418 (34·4) | 4391 (12·8) | 18 (4·1) |
| AHF | 1477 (9·4) | 263 (0·8) | 180 (41·0) |
| Stroke | 1265 (8·0) | 1282 (3·8) | 62 (14·1) |
| Malignancy | 1413 (9·0) | 808 (2·4) | 234 (53·3) |
| **Severity of illness** | | | |
| APSIII | 42·00 [33·00, 55·00] | 42·00 [29·00, 59·00] | 78·00 [55·00, 91·00] |
| OASIS | 34·00 [28·00, 40·00] | 32·00 [26·00, 40·00] | 31·00 [25·00, 35·00] |
| SOFA | 4·00 [2·00, 6·00] | 5·00 [4·00, 8·00] | 15·00 [12·00, 17·00] |
| MODS | 5·00 [2·00, 7·00] | 3·00 [2·00, 6·00] | 13·00 [11·00, 14·00] |
| SAPS | 20·00 [17·00, 23·00] | 20·00 [17·00, 24·00] | 22·00 [18·00, 24·00] |
| **Outcomes** | | | |
| Days of hospital admission (d), (median, IQR) | 7·88 [5·05, 12·81] | 6·32 [3·96, 10·04] | 10·71 [7·06, 16·91] |
| Days of ICU admission (d), (median, IQR) | 2·75 [1·71, 4·99] | 2·53 [1·69, 4·14] | 3·43 [2·11, 6·72] |
| Hospital mortality, n (%) | 2353 (14·89) | 3966 (11·60) | 51 (11·62) |

*BMI body mass index, CCU coronary care unit, CSRU cardiac surgery recovery unit, MICU medical ICU, SICU surgical ICU, TSICU trauma/surgical ICU.*

**Development of prediction model**

According to Figure 1, three prediction models were obtained from different data sources. The performance of models was evaluated by the aforementioned parameters (AUC, sensitivity, specificity, F1 and accuracy) and internal/ external validation. 'Model 3' had the best performance using

MIMIC-III and eICU-CRD cohorts, which presents universality and robustness in different hospitals and centers. 'Model 2' is acquired employing multicenter datasets of the eICU-CRD database, and the performance of which is not as good as 'Model1'.

**Table 3: Summary of the optimal model's cross validation performance in multicenter databases**

| Indexes | Training set / Testing set | MIMIC-III | eICU-CRD | MIMIC-III - eICU-CRD |
|---|---|---|---|---|
| **AUC** | MIMIC-III | 0·866 | 0·823 | 0·858 |
| | eICU-CRD | 0·837 | 0·849 | 0·849 |
| | PLAGH-S | 0·835 | 0·811 | 0·838 |
| **Sensitivity** | MIMIC-III | 0·802 | 0·774 | 0·834 |
| | eICU-CRD | 0·752 | 0·841 | 0·763 |
| | PLAGH-S | 0·725 | 0·627 | 0·882 |
| **Specificity** | MIMIC-III | 0·760 | 0·712 | 0·705 |
| | eICU-CRD | 0·772 | 0·700 | 0·784 |
| | PLAGH-S | 0·820 | 0·858 | 0·691 |
| **F1** | MIMIC-III | 0·681 | 0·624 | 0·664 |
| | eICU-CRD | 0·676 | 0·655 | 0·652 |
| | PLAGH-S | 0·690 | 0·722 | 0·701 |
| **Accuracy** | MIMIC-III | 0·876 | 0·871 | 0·875 |
| | eICU-CRD | 0·891 | 0·901 | 0·901 |
| | PLAGH-S | 0·882 | 0·900 | 0·882 |

*AUROC (AUC) Area under the receiver operating characteristic curves, MIMIC-III - eICU-CRD the training set with the fusion of MIMIC-III's cohort and eICU-CRD's cohort*

**Explanation of risk factors**

In order to improve the clinical significance of the model, the following results were obtained: key risk factors affecting the outcome of elderly patients with MODS and how the factors' value was related to the patient outcome; how to visually explain the early predictive risk of a single patient, which would help doctors understand the analytic process of the prediction model. We adopt 'Model3' as our optimal early risk prediction model to assess patient outcome.

Figure 3(a) displays the top 20 risk factors in our model. The features ranking (y axis) implies the importance of the prediction model. The SHAP value (x axis) is a unified index responding to the impact of a feature in the model. In each feature importance row, all patients' attribution to outcome were plotted using different color dots, in which the red (blue) dot represented high (low) value. Figure 3(b) shows the top 20 most important features evaluated by the average of absolute SHAP value. Risk factors such as GCS (max, avg and min), Respiratory Rate (rr: avg), Blood Urea Nitrogen (BUN: min), age, shock index (si: avg), SIRS score (avg), total urine output of 24 hours (uo) and Blood Oxygen Saturation ($SpO_2$: min) during the initial 24 hours of ICU stay were ranked as the ten most important factors. The most important feature of GCS is to emphasize the nervous system as the most crucial for early assessment of patient's status than other organ systems. BUN level was found to be a key factor, which is easily overlooked by clinicians. Figure 3(c) provides two typical relative samples to illustrate the interpretability of the model. Although the patient had normal urine output (2825 ml), BMI (31·4 kg/m$^2$) and $SpO_2$ (98·76 %), he died due to poor GCS (7 points), high BUN (59 mg/dL) and $FiO_2$ (60 %). While the survivor had normal GCS (15 points), SIRS score (1 point), BUN (9 mg/dL) and no mechanical ventilation need. Although at very high age (87 years old) and had high blood glucose level (148 mg/dL), the patient survived after treatment. The risk factor rankings of 'Model 1' and 'Model 2' can be found in the supplementary files (eFigure 5 and eFigure 6). A summary of three models' risk factors ranking was presented in eTable 6. With different training sets, the feature importance rankings remained largely unchanged, which demonstrated the robustness and stability of the SHAP value employment to assess risk factors.

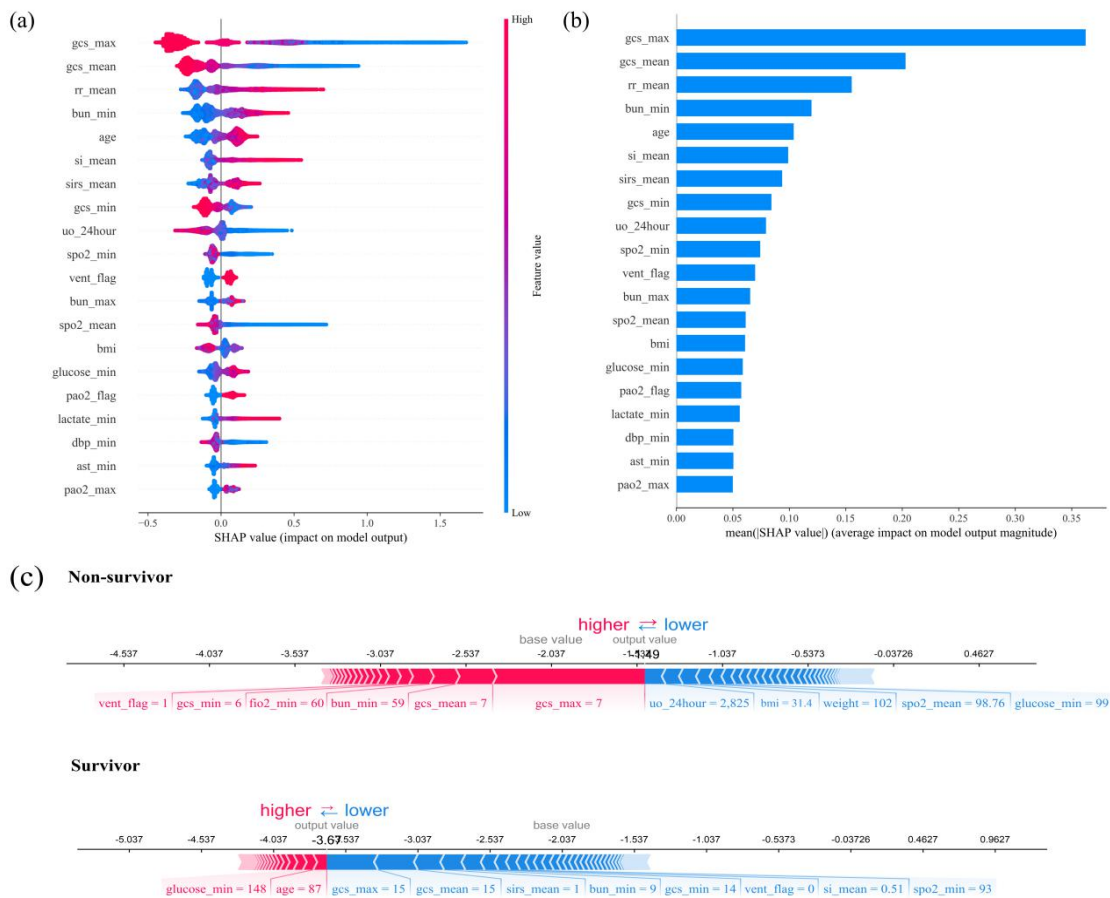

**Figure 3: The model's interpretation. (a). The importance ranking of the top 20 risk factors with stability and interpretation using the optimal model. The higher SHAP value of a feature is given, the higher risk of death the patient would have. The red part in feature value represents higher value; (b). The importance ranking of the top 20 variables according to the mean (|SHAP value|); (c). The interpretation of model prediction results with the two samples**

**Predictive performance evaluation**

The model's performance validated by using the MIMIC-III cohort is presented in Figure 4 and summarized in eTable 7. Consistently, our model (AUC with 95% confidence interval (CI): 0·858 [0·841 - 0·875]) was superior to 5 baseline machine learning models and commonly used clinical scores. The model's performance evaluated with the eICU-CRD and PLAGH-S cohorts is presented in the supplementary files (eFigure 8, eTable 8, eFigure 9 and eTable 9). Although the AUC of the our novel model was slightly lower than that of SVM (eFigure 9), the sensitivity of the model was significantly better. These results have demonstrated satisfactory predictive ability and universality of the model. We retained the top 20 features as input values to evaluate the model's performance when multiple features required by the model were not measured or documented.

As shown in eFigure 10, good prediction performance was achieved in three validation datasets (AUC 0·858 of 130 features vs AUC 0·851 of 20 features using the MIMIC-III cohort, AUC 0·849 of 130 features vs AUC 0·839 of 20 features using the eICU-CRD cohort, AUC 0·838 of 130 features vs AUC 0·776 of 20 features using the PLAGH-S cohort). Consistently, all results of this model outperformed all clinical scores listed in eTable 10. The model's performance when validated leveraging the total cohort from MIMIC-III and eICU-CRD was presented in eTable 11, which provides extra information for further evaluation of the model's performance.

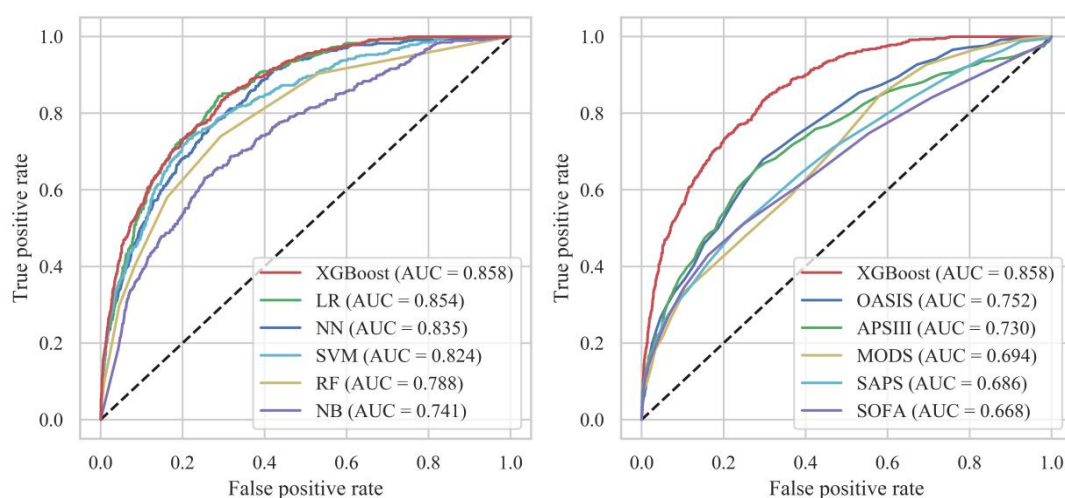

**Figure 4: Our model's performance, evaluated in the MIMIC-III cohort, comparing with the baseline models and clinical scores**

## Discussion

In this study, a generalizable and interpretable model for the early prediction of mortality risk in elderly patients with MODS was developed and validated based on three clinical databases, and performed well in the early assessment of organ damage, providing crucial support for clinical decision making. Generally, four part of contributions were accomplished in this study: (1) Comprehensive and reasonable experiments were designed to establish an optimal prediction model with satisfactory performance, utilizing training sets from both single center with long time span and multiple centers with short time span as well as the fusion of both; (2) Multiple methods and indexes were utilized to evaluate the model's performance, including internal and external validations covering the countries with different levels of economic development. Meanwhile, comparisons with machine learning methods and clinical scores were also included. The indexes adopted were AUC, sensitivity, specificity, F1, accuracy, robustness and universality; (3) The explanation was employed to the black box tree ensemble model of XGBoost, which helps doctors better understand the decision-making process of the model; (4) Key risk factors with robustness and interpretation were acquired. Two important factors of BUN and shock index, significantly correlated to death, also were discovered, which were commonly overlooked by physicians and commonly used clinical scores.

Most previous researchers examined the early mortality prediction performance of the proposed models by cross validation limited to their single-center study dataset. Although their models performed well in that setting, their performances in other datasets were not verified. In this study, the model proposed by us was evaluated using large sample-size and multicenter databases, which enabled us to accomplish the internal and external validations without overlapping. To the best of our knowledge, this is the first study in which mortality prediction and assessment models were validated in countries with different economic development levels (US & China). In order to comprehensively evaluate the model's performance, the commonly used clinical scores and machine learning methods were both included and five measurements were adopted to compare both of them. Moreover, the new score of OASIS, which had been proven to have more concise inputs and better performance compared to the APACHE score was also included for comparison.[18,29] Finally, our model consistently outperformed all the baseline models and clinical scores (Figure 4, eFigure 8 and eFigure 9). Thus, the model bears great potential to be generalized and applied in clinical practice.

Several recent studies have reported their models' validation performance using multicenter datasets. Matthew MC et al. developed a model using 60% of the data from a multicenter database and evaluated it using the remaining 40% of the data.[29] Shamin N et al. utilized a regional clinical database to develop the model and externally validated it utilizing a database from another region.[30] Benjamin S et al. adopted their local database to train the model and assessed it with combined internal and external validations.[19] Hamid M et al. leveraged three institutional databases and trained one model from each institution which was evaluated using databases from the remaining institutions.[31] However, these studies did not specify an approach to get a generalizable and robust model. Thus, we chose the special validation sets where no overlapping exists among training sets to evaluate the models' performance (accuracy and robustness). The validation sets in our study included 20% of the MIMIC-III cohort, 20% of the eICU-CRD cohort and total cohort of the PLAGH-S. Data used for training were the remaining data from each database, namely MIMIC-III and eICU-CRD, as well as the combination of them. Therefore, three different types of training sets (single-center with long duration, multicenter with short duration and the fusion of them) were analyzed and validated, which enabled us to develop this optimal predictive model.

Recently, interest in using the interpretation and tree ensemble models have been growing to develop mortality prediction model, such as RF and Gradient Boosting Decision Tree (GBDT).[29,32,33] Although tree ensemble models are more accurate compared to LR and can provide features importance ranking, it cannot tell clinicians whether the important factors are protective or dangerous like LR dose: the 'black-box' nature of machine learning algorithms can make it difficult to understand and correct errors when they occur.[34] Meanwhile, a trade-off between the accuracy and interpretation of the models is often hard to achieve. In addition, the risk probability output of the model is not easily understandable for doctors. Therefore, we applied the proposed SHAP values to the XGBoost, which made our model able to achieve both optimal accuracy and interpretability to provide more insights to doctors. Detailed information was described in results, as well as explanation of risk factors. In summary, given the key risk factors, the model can visually explain to clinicians which specific features of elderly patients with MODS predispose them to high (low) risk of death.

Most of the current predictive models based on tree (ensemble) methods generate features important rankings using default parameters of 'importance type'. Actually, there are five alternative types available (ie., 'weight', 'gain', 'cover', 'total_gain' and 'total_cover') and the ranking results would be changed when different types were chosen. Top 30 features ranking results in three commonly used types were illustrated in eFigure 14. and eTable 8, which intuitively presented its inconsistency. Different from the previous studies, we adopted the SHAP value, a unified measure, while generating features important ranking. It had been theoretically proven to be an optimal approach and the only possible consistent feature attribution method.[28] Its consistency has been proved with the similar feature ranking even though using the multiple sources of training data in as shown in eTable 9.

The key risk factors identified in this study are consistent with clinicians' prior knowledge. Two statistically significant factors identified, BUN and shock index, ranked fourth and sixth, respectively, were usually overlooked by physicians and contributed in mortality assessment in elderly with MODS. BUN is a laboratory test used to evaluate renal function, a high value of which might be related to renal failure, hypovolemia, congestive heart failure and increased catabolism, etc. However, most of the clinical scores (like SOFA, APACHE-II) only include creatinine level. Recently, several studies found BUN to be an independent factor (biomarker) for mortality in ICU patients, and that it can also indicate the degree of heart failure by reflecting the interaction of nutrition, protein metabolism and renal status.[35,36] Okan A et al. found that high BUN level to be indicative of ongoing multi-organ failure.[36] While these studies are limited by small sample sets and utilization of early simplified models, results of our study were based on the large sample sets and sophisticated models. Ryan W. H et al. recently reported that persistent elevation of BUN indicates concurrent muscle bioenergetic failure, muscle catabolism/altered protein homeostasis and persistent muscle loss, which further affects the metabolic process and aggravates disease severity.[37] Shock index was generally utilized to measure the severity of sepsis and septic shock.[38] Some studies also demonstrate it to be a mortality predictor or indicator for certain kinds of diseases.[39,40] As for elderly patients suffering from multiple organ dysfunction, our results indicate that shock index could be used as an important factor to predict patient death. Before the study, we speculated that a large proportion of the patients in the dataset would have sepsis or septic shock. Consistently, the sepsis population ratio and its death rate for each study cohort were relatively high (47·08% of sepsis and 8·53% of mortality in the MIMIC-III cohort, 63·60% of sepsis and 9·47% of mortality in the MIMIC-III cohort, and 59·68% of sepsis and 8·89% of mortality in the MIMIC-III cohort). A more in-depth investigation of the disease mechanism warrants further studies.

Interestingly, two risk factors, GCS (nervous system) and respiratory rate (respiratory system), ranking No. 1 and 3 respectively, were found to be critical in assessing disease severity, which is consistent to the conclusion of a previous study.[19] This finding indicates that the state of the nervous and respiratory systems should be paid more attention to, rather than being treated equally as other systems. For a more comprehensive assessment of our model's robustness, we selected the top 20 features acquired by SHAP values to train the model, which were evaluated in three study cohorts. Consequently, our model achieved satisfactory performance which outperforms all the commonly used clinical scores.

This study has some limitations. Although the predictive model in this study achieved early mortality prediction in elderly patients with MODS, it would be of greater value to accomplish a real-time evaluation of disease severity. Therefore, we plan to leverage the time series models such as long short-term memory (LSTM) to develop real-time prediction models. Moreover, the PLAGH-S database was constructed to facilitate the management of surgical ICU data in our hospital, which is a single ICU dataset compared to the MIMIC-III and eICU-CRD database where all types of ICU wards are included. We evaluated the model using the PLAGH-S' cohort, which failed to accurately reflect the performance of the model. Therefore, we will further complete data standardization and sort out the cohort from the PLAGH General Intensive Care Unit Database, which contains the hospitalization records of above 66,227 adult patients in nine ICUs in the past ten years. This will allow us to provide readers with more accurate multicenter validation results in different countries.

In conclusion, based on the multicenter clinical databases originating from different regions, we established a generalized and interpretable predictive model with optimal performance utilizing the fusion data of the MIMIC-III and eICU-CRD cohorts for early evaluation of mortality risk in elderly with MODS. The top ten risk factors with the greatest predictive value are GCS, respiratory rate, BUN, age, shock index, SIRS score, total urine output on the first day in ICU and $SpO_2$. Meanwhile, BUN and shock index are factors worth more attention from clinicians for their predictive value of mortality in elderly patients with organ dysfunction.

**Code and data available**

The code that was used to extract code from the MIMIC-III and eICU-CRD databases, develop machine learning models and calculate statistical analysis and part of source data are available at https://github.com/liuxiaoliXRZS/MODSE.

**Interpretable Machine Learning Model for Early Prediction of Mortality in Elderly Patients with Multiple Organ Dysfunction Syndrome (MODS): a Multicenter Retrospective Study and Cross Validation**

Online Supplement

**Additional Methods**
**Additional Reference**

eFigure 1: Study cohort in the MIMIC-III database. Detailed inclusion with the number of cohorts were indicated

eFigure 2: Study cohort in the eICU-CRD database. Detailed inclusion with the number of cohorts were indicated

eFigure 3: Study cohort in the PLAGH-S database. Detailed inclusion with the number of cohorts were indicated

eFigure 4: The importance ranking of top 30 risk factors utilizing the custom methods

eFigure 5: The importance ranking of the top 20 risk factors with stability and interpretation employing the MIMIC-III cohort

eFigure 6: The importance ranking of the top 20 risk factors with stability and interpretation employing the eICU-CRD cohort

eFigure 7: The relationship between the risk factors and death with the top 6. In each subgraph, the most correlation feature with this risk factor also be presented

eFigure 8: Our model's performance, evaluated in the eICU-CRD cohort, comparing with the baseline models and clinical scores

eFigure 9: Our model's performance, evaluated in the PLAGH-S cohort, comparing with the baseline models and clinical scores

eFigure 10: The cross-validation of our model with the top 20 importance risk factors in the multi-center databases

eTable 1: Baseline characteristic of the patients be included in the MIMIC-III cohort

eTable 2: Baseline characteristic of the patients be included in the eICU-CRD cohort

eTable 3: Baseline characteristic of the patients be included in the PLAGH-S cohort

eTable 4: The features' missing ratio before being processed

eTable 5: The variables information of top 30 risk factors utilizing the custom methods

eTable 6: The variables information of top 20 risk factors with prediction models developing by different databases

eTable 7: The detailed information of the optimal model's performance evaluated in the MIMIC-III cohort

eTable 8: The detailed information of the optimal model's performance evaluated in the eICU-CRD cohort

eTable 9: The detailed information of the optimal model's performance evaluated in the PLAGH-S cohort

eTable 10: The detailed information of our model's performance with cross-validation in the multicenter databases

eTable 11: The additional information of cross-validation leveraging the total cohorts as the testing set

## Additional Methods
### Data extraction
Data was separately extracted and limited to the first 24h for patients staying in ICU from three databases adopting Postgresql language. The static information (ie. demographic) were directly extracted and named according to their meaning. The dynamic information (ie. vital signs, laboratory measurement, output and clinical treatment) were totally extracted. Due to the 208 critical care units included in eICU-CRD database, the critical care units' name and admission type of patients were inconsistently recorded. Therefore, the types of ICU were categorized as 'CSRU', 'SICU', 'CCU', 'MICU', 'NICU' or 'other/Unknown', and the types of admission were categorized as 'URGENT', 'EMERENCY', 'FLOOR', 'Intermediate Care Unit' or 'Other/Unknown'. Moreover, only the clinical score of APS-III was recorded in eICU-CRD database. Therefore, the other clinical scores were calculated for each study cohort according to the definition.

### Feature selection
Before further calculating the statistical features, we dropped the outliers of them using the range of features that have been obtained. Firstly, the initial values expected base excess were received logarithmic transformation and the interquartile range (IQR) method was adopted to get the lower and upper bound which should be further calculated to get the inverse of logarithm bounds. Then, the clinician helped us listing the physiological boundaries of each feature. Finally, both considering the two ways, we acquired the final utilizing boundaries. The missing values of the feature were imputed using the median value. While the missing values of $FiO_2$ were imputed to 21. Moreover, the feature missing ratio exceeding 30% was added additional information to indicate whether it was being measured. And we used 'flag' (with the imputation of '0' or '1') to represent them indicating whether it be or not be recorded. In order to uniformly each of feature's name in three study cohorts, we called them using the features' abbreviation with the add of statistical or flag name, like 'hr_min', 'lactate_max' and 'dobutamine_flag', which was convenient for subsequent comparative analysis.

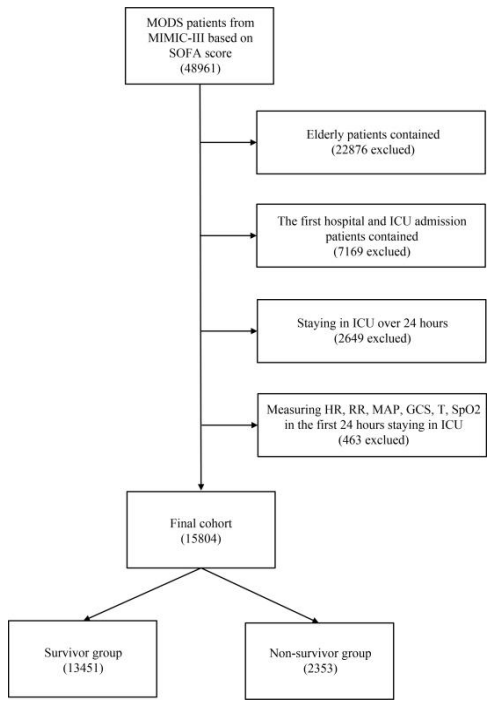

**eFigure 1: Study cohort in the MIMIC-III database. Detailed inclusion with the number of cohorts were indicated**

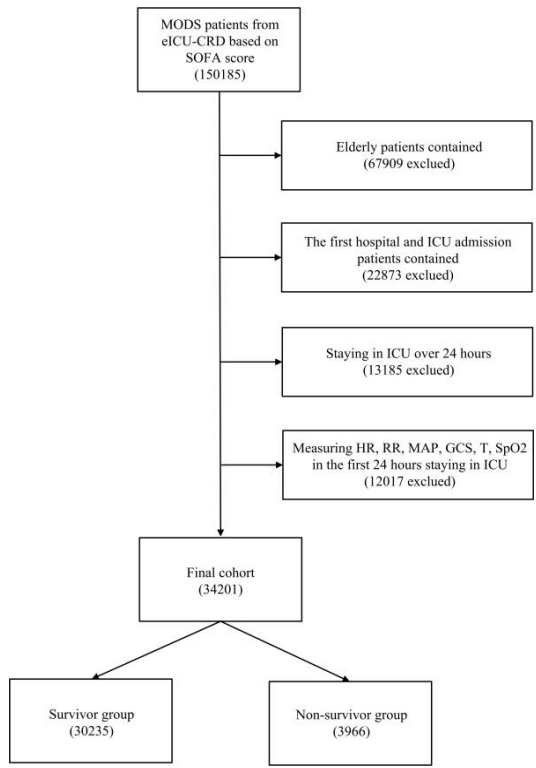

**eFigure 2: Study cohort in the eICU-CRD database. Detailed inclusion with the number of cohorts were indicated**

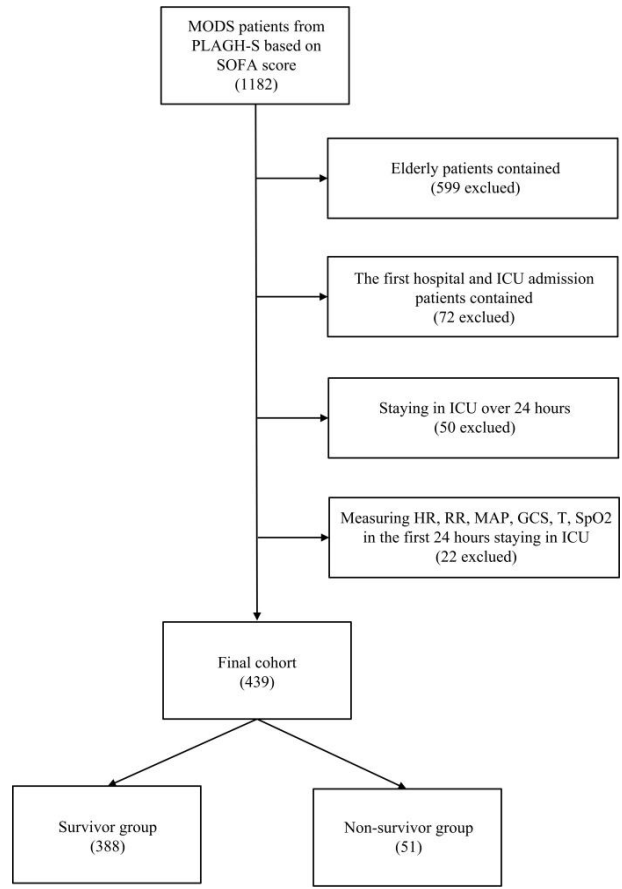

eFigure 3: Study cohort in the PLAGH-S database. Detailed inclusion with the number of cohorts were indicated

eTable 1: Baseline characteristic of the patients be included in the MIMIC-III cohort

|  | All Patients (n = 15804) | Survivors (n = 13451) | Non-survivors (n = 2353) | p |
|---|---|---|---|---|
| **Demographic** | | | | |
| Age (yr), (median, IQR) | 77·00 [71·00, 83·00] | 77·00 [71·00, 83·00] | 79·00 [73·00, 85·00] | <0·001 |
| Male, n (%) | 8281 (52·4) | 7108 (52·8) | 1173 (49·9) | 0·008 |
| Weight (kg), (median, IQR) | 77·20 [65·00, 90·70] | 78·10 [65·80, 91·40] | 73·10 [61·00, 86·00] | <0·001 |
| Height (cm), (median, IQR) | 167·64 [160·02, 175·26] | 167·64 [160·02, 175·26] | 166·37 [160·00, 175·26] | 0·006 |
| BMI (kg/m$^2$), (median, IQR) | 27·97 [24·37, 32·07] | 28·11 [24·58, 32·22] | 26·87 [23·08, 31·18] | <0·001 |
| ICU type (%) | | | | <0·001 |
| CCU | 2738 (17·3) | 2373 (17·6) | 365 (15·5) | |
| CSRU | 3509 (22·2) | 3331 (24·8) | 178 (7·6) | |
| MICU | 5428 (34·3) | 4373 (32·5) | 1055 (44·8) | |
| SICU | 2473 (15·6) | 2014 (15·0) | 459 (19·5) | |
| TSICU | 1656 (10·5) | 1360 (10·1) | 296 (12·6) | |
| Ethnicity (%) | | | | <0·001 |
| ASIAN | 360 (2·3) | 300 (2·2) | 60 (2·5) | |
| BLACK | 903 (5·7) | 800 (5·9) | 103 (4·4) | |
| HISPANIC | 264 (1·7) | 227 (1·7) | 37 (1·6) | |
| OTHER | 2504 (15·8) | 2014 (15·0) | 490 (20·8) | |
| WHITE | 11773 (74·5) | 10110 (75·2) | 1663 (70·7) | |
| Admission type (%) | | | | <0·001 |
| ELECTIVE | 2332 (14·8) | 2244 (16·7) | 88 (3·7) | |
| EMERGENCY | 13006 (82·3) | 10808 (80·4) | 2198 (93·4) | |
| URGENT | 466 (2·9) | 399 (3·0) | 67 (2·8) | |
| **Comorbidities** | | | | |
| CLF | 436 ( 2·8) | 317 ( 2·4) | 119 (5·1) | <0·001 |
| ALF | 250 ( 1·6) | 115 ( 0·9) | 135 (5·8) | <0·001 |
| COPD | 3542 (22·5) | 2964 (22·1) | 578 ( 24·8) | 0·005 |
| ARDS | 2960 (18·8) | 1925 (14·4) | 1035 ( 44·3) | <0·001 |
| CAD | 7129 (45·3) | 6330 (47·2) | 799 ( 34·2) | <0·001 |
| CRF | 2248 (14·3) | 1902 (14·2) | 346 ( 14·8) | 0·446 |
| AKI | 4165 (26·5) | 3085 (23·0) | 1080 ( 46·3) | <0·001 |
| CHF | 5418 (34·4) | 4515 (33·7) | 903 (38·7) | <0·001 |
| AHF | 1477 ( 9·4) | 1280 ( 9·6) | 197 (8·4) | 0·095 |
| Stroke | 1265 ( 8·0) | 857 ( 6·4) | 408 (17·5) | <0·001 |
| Malignancy | 1413 ( 9·0) | 1051 ( 7·8) | 362 (15·5) | <0·001 |
| **Severity of illness** | | | | |
| APSIII | 42·00 [33·00, 55·00] | 41·00 [32·00, 51·00] | 57·00 [43·00, 75·00] | <0·001 |
| OASIS | 34·00 [28·00, 40·00] | 33·00 [28·00, 38·00] | 40·00 [35·00, 47·00] | <0·001 |
| SOFA | 4·00 [2·00, 6·00] | 4·00 [2·00, 6·00] | 6·00 [3·00, 9·00] | <0·001 |
| MODS | 5·00 [2·00, 7·00] | 4·00 [2·00, 7·00] | 6·00 [4·00, 10·00] | <0·001 |
| SAPS | 20·00 [17·00, 23·00] | 19·00 [16·00, 22·00] | 22·00 [19·00, 26·00] | <0·001 |
| **Outcomes** | | | | |
| Days of hospital admission (d), (median, IQR) | 7·88 [5·05, 12·81] | 7·94 [5·24, 12·73] | 7·01 [3·28, 13·66] | <0·001 |
| Days of ICU admission (d), (median, IQR) | 2·75 [1·71, 4·99] | 2·57 [1·66, 4·50] | 4·17 [2·06, 8·19] | <0·001 |

**eTable 2: Baseline characteristic of the patients be included in the eICU-CRD cohort**

| | All Patients (n = 34201) | Survivors (n = 30235) | Non-survivors (n = 3966) | p |
|---|---|---|---|---|
| **Demographic** | | | | |
| Age (yr), (median, IQR) | 76·00 [70·00, 83·00] | 76·00 [70·00, 83·00] | 78·00 [71·00, 84·00] | <0·001 |
| Male, n (%) | 17586 (51·4) | 15551 (51·4) | 2035 (51·3) | 0·898 |
| Weight (kg), (median, IQR) | 77·30 [64·40, 92·00] | 77·60 [64·90, 92·30] | 75·10 [62·10, 89·90] | <0·001 |
| Height (cm), (median, IQR) | 167·60 [160·00, 177·00] | 167·60 [160·00, 177·00] | 167·60 [160·00, 175·30] | 0·414 |
| BMI (kg/m$^2$), (median, IQR) | 27·10 [23·30, 31·80] | 27·20 [23·40, 31·90] | 26·40 [22·60, 31·20] | <0·001 |
| ICU type (%) | | | | <0·001 |
| CCU | 7077 ( 20·7) | 6380 ( 21·1) | 697 ( 17·6) | |
| CSRU | 902 (2·6) | 769 (2·5) | 133 (3·4) | |
| MICU | 21382 (62·5) | 18721 ( 61·9) | 2661 ( 67·1) | |
| NICU | 2860 (8·4) | 2573 (8·5) | 287 (7·2) | |
| SICU | 1980 (5·8) | 1792 (5·9) | 188 (4·7) | |
| Ethnicity (%) | | | | 0·25 |
| ASIAN | 386 (1·1) | 349 (1·2) | 37 (0·9) | |
| BLACK | 2850 (8·3) | 2539 (8·4) | 311 (7·8) | |
| HISPANIC | 1259 (3·7) | 1128 (3·7) | 131 (3·3) | |
| OTHER/UNKNOWN | 1541 (4·5) | 1366 (4·5) | 175 (4·4) | |
| WHITE | 28165 (82·4) | 24853 (82·2) | 3312 (83·5) | |
| Admission type (%) | | | | <0·001 |
| ELECTIVE | 12974 (37·9) | 11342 (37·5) | 1632 (41·1) | |
| FLOOR | 3400 (9·9) | 2849 (9·4) | 551 (13·9) | |
| INTERMEDIATE CARE UNIT | 2347 (6·9) | 2066 (6·8) | 281 (7·1) | |
| OTHER/KNOWN | 9627 (28·1) | 8589 (28·4) | 1038 (26·2) | |
| URGENT | 5853 (17·1) | 5389 (17·8) | 464 (11·7) | |
| **Comorbidities** | | | | |
| CLF | 225 (0·7) | 174 (0·6) | 51 (1·3) | <0·001 |
| ALF | 58 (0·2) | 38 (0·1) | 20 (0·5) | <0·001 |
| COPD | 4250 (12·4) | 3720 (12·3) | 530 (13·4) | 0·063 |
| ARDS | 7655 (22·4) | 5923 (19·6) | 1732 (43·7) | <0·001 |
| CAD | 3069 (9·0) | 2684 (8·9) | 385 (9·7) | 0·094 |
| CRF | 3635 (10·6) | 3104 (10·3) | 531 (13·4) | <0·001 |
| AKI | 4514 (13·2) | 3528 (11·7) | 986 (24·9) | <0·001 |
| CHF | 4391 (12·8) | 3781 (12·5) | 610 (15·4) | <0·001 |
| AHF | 263 (0·8) | 217 (0·7) | 46 (1·2) | 0·004 |
| Stroke | 1282 (3·8) | 1087 (3·6) | 195 (4·9) | <0·001 |
| Malignancy | 808 (2·4) | 664 (2·2) | 144 (3·6) | <0·001 |
| **Severity of illness** | | | | |
| APSIII | 42·00 [29·00, 59·00] | 41·00 [28·00, 55·00] | 64·00 [45·00, 89·00] | <0·001 |
| OASIS | 32·00 [26·00, 40·00] | 31·00 [25·00, 38·00] | 41·00 [34·00, 49·00] | <0·001 |
| SOFA | 5·00 [4·00, 8·00] | 5·00 [4·00, 7·00] | 8·00 [6·00, 11·00] | <0·001 |
| MODS | 3·00 [2·00, 6·00] | 3·00 [2·00, 5·00] | 7·00 [4·00, 10·00] | <0·001 |
| SAPS | 20·00 [17·00, 24·00] | 20·00 [17·00, 23·00] | 25·00 [21·00, 30·00] | <0·001 |
| **Outcomes** | | | | |
| Days of hospital admission (d), (median, IQR) | 6·32 [3·96, 10·04] | 6·44 [4·08, 10·06] | 5·25 [2·77, 9·78] | <0·001 |
| Days of ICU admission (d), (median, IQR) | 2·53 [1·69, 4·14] | 2·43 [1·67, 3·98] | 3·35 [1·89, 6·46] | <0·001 |

### eTable 3: Baseline characteristic of the patients be included in the PLAGH-S cohort

| | All Patients (n = 439) | Survivors (n = 388) | Non-survivors (n = 51) | p |
|---|---|---|---|---|
| **Demographic** | | | | |
| Age (yr), (median, IQR) | 73·00 [68·00, 79·50] | 73·00 [68·00, 79·00] | 73·00 [66·50, 81·50] | 0·911 |
| Male, n (%) | 248 (56·5) | 220 (56·7) | 28 (54·9) | 0·926 |
| Weight (kg), (median, IQR) | 65·00 [58·00, 72·00] | 65·00 [60·00, 73·00] | 60·00 [53·00, 69·00] | 0·006 |
| Height (cm), (median, IQR) | 165·00 [160·00, 172·00] | 165·00 [160·00, 172·00] | 166·00 [159·50, 172·00] | 0·868 |
| BMI (kg/m$^2$), (median, IQR) | 24·00 [22·00, 27·00] | 24·00 [22·00, 27·00] | 23·00 [20·00, 26·00] | 0·02 |
| ICU type (%) | | | | |
| SICU | 439 (100·0) | 388 (100·0) | 51 (100·0) | |
| Ethnicity (%) | | | | |
| ASIAN | 439 (100·0) | 388 (100·0) | 51 (100·0) | |
| Admission type (%) | | | | 0·006 |
| EMERGENCY | 141 (32·1) | 122 (31·4) | 19 (37·3) | |
| OTHER/UNKNOWN | 259 (59·0) | 237 (61·1) | 22 (43·1) | |
| OUTPATIENT | 39 (8·9) | 29 (7·5) | 10 (19·6) | |
| **Comorbidities** | | | | |
| CLF | 8 ( 1·8) | 7 ( 1·8) | 1 (2·0) | 1 |
| ALF | 77 (17·5) | 56 (14·4) | 21 ( 41·2) | <0·001 |
| COPD | 37 ( 8·4) | 23 ( 5·9) | 14 ( 27·5) | <0·001 |
| ARDS | 66 (15·0) | 42 (10·8) | 24 ( 47·1) | <0·001 |
| CAD | 181 (41·2) | 160 (41·2) | 21 (41·2) | 1 |
| CRF | 31 ( 7·1) | 22 ( 5·7) | 9 ( 17·6) | 0·004 |
| AKI | 77 (17·5) | 55 (14·2) | 22 (43·1) | <0·001 |
| CHF | 18 ( 4·1) | 16 ( 4·1) | 2 (3·9) | 1 |
| AHF | 180 (41·0) | 136 (35·1) | 44 ( 86·3) | <0·001 |
| Stroke | 62 (14·1) | 49 (12·6) | 13 (25·5) | 0·023 |
| Malignancy | 234 (53·3) | 209 (53·9) | 25 ( 49·0) | 0·615 |
| **Severity of illness** | | | | |
| APSIII | 78·00 [55·00, 91·00] | 77·00 [53·75, 91·00] | 88·00 [75·50, 106·50] | <0·001 |
| OASIS | 31·00 [25·00, 35·00] | 32·00 [25·00, 35·00] | 31·00 [27·50, 36·00] | 0·554 |
| SOFA | 15·00 [12·00, 17·00] | 14·00 [12·00, 16·00] | 18·00 [14·00, 19·00] | <0·001 |
| MODS | 13·00 [11·00, 14·00] | 13·00 [11·00, 14·00] | 15·00 [11·00, 17·00] | 0·002 |
| **Outcomes** | | | | |
| Days of hospital admission (d), (median, IQR) | 10·71 [7·06, 16·91] | 10·13 [6·86, 15·08] | 20·98 [8·92, 37·29] | <0·001 |
| Days of ICU admission (d), (median, IQR) | 3·43 [2·11, 6·72] | 3·15 [2·03, 5·90] | 7·28 [4·21, 20·98] | <0·001 |

### eTable 4: The features' missing ratio before being processed

| Features' name | MIMIC-III (%) | eICU-CRD (%) | PLAGH-S (%) |
|---|---|---|---|
| **Demographic** | | | |
| age | 0 | 0 | 0 |
| bmi | 52·73 | 1·14 | 7·97 |
| gender | 0 | 0 | 0 |
| height | 36·08 | 1·13 | 0 |
| weight | 29·35 | 0·39 | 7·97 |

| Laboratory test | | | |
|---|---|---|---|
| albumin_max | 64·57 | 40·22 | 0·23 |
| albumin_min | 64·57 | 40·22 | 0·23 |
| alkaline_phosphatase_max | 58·96 | 43·69 | 0·23 |
| alkaline_phosphatase_min | 58·96 | 43·69 | 0·23 |
| alt_max | 58·03 | 42·93 | 0·23 |
| alt_min | 58·03 | 42·93 | 0·23 |
| ast_max | 58·09 | 41·97 | 0·46 |
| ast_min | 58·09 | 41·97 | 0·46 |
| be_max | 32·3 | 61·97 | 0 |
| be_min | 32·3 | 61·97 | 0 |
| bicarbonate_max | 0·8 | 9·25 | 0 |
| bicarbonate_min | 0·8 | 9·25 | 0 |
| bilirubin_max | 58·49 | 43·68 | 0·23 |
| bilirubin_min | 58·49 | 43·68 | 0·23 |
| bnp_max | 94·99 | 85·1 | 0·46 |
| bnp_min | 94·99 | 85·1 | 0·46 |
| bun_max | 0·28 | 3·58 | 0·23 |
| bun_min | 0·28 | 3·58 | 0·23 |
| chloride_max | 0·58 | 3·66 | 0·23 |
| chloride_min | 0·58 | 3·66 | 0·23 |
| creatinine_max | 0·24 | 3·19 | 0·23 |
| creatinine_min | 0·24 | 3·19 | 0·23 |
| fibrinogen_max | 75·15 | 91·8 | 0·46 |
| fibrinogen_min | 75·15 | 91·8 | 0·46 |
| glucose_max | 0·27 | 1·48 | 0 |
| glucose_min | 0·27 | 1·48 | 0 |
| hematocrit_max | 0·28 | 3·36 | 0·23 |
| hematocrit_min | 0·28 | 3·36 | 0·23 |
| hemoglobin_max | 0·49 | 3·62 | 0 |
| hemoglobin_min | 0·49 | 3·62 | 0 |
| inr_max | 8·15 | 41·06 | 0·46 |
| inr_min | 8·15 | 41·06 | 0·46 |
| lactate_max | 40·56 | 60·88 | 0 |
| lactate_min | 40·56 | 60·88 | 0 |
| lymphocytes_max | 49·73 | 30·76 | 0·23 |
| lymphocytes_min | 49·73 | 30·76 | 0·23 |
| magnesium_max | 5·16 | 38·27 | 0·23 |
| magnesium_min | 5·16 | 38·27 | 0·23 |
| neutrophils_max | 49·51 | 37·48 | 0·23 |
| neutrophils_min | 49·51 | 37·48 | 0·23 |
| paco2_max | 32·3 | 52·57 | 0 |
| paco2_min | 32·3 | 52·57 | 0 |
| pafi_max | 49·27 | 62·69 | 0 |
| pafi_min | 49·27 | 62·69 | 0 |
| pao2_max | 32·3 | 52·29 | 0 |
| pao2_min | 32·3 | 52·29 | 0 |
| ph_max | 32·3 | 53·02 | 0 |
| ph_min | 32·3 | 53·02 | 0 |
| platelet_max | 0·61 | 4·3 | 0·23 |
| platelet_min | 0·61 | 4·3 | 0·23 |
| potassium_max | 0·18 | 2·67 | 0·23 |
| potassium_min | 0·18 | 2·67 | 0·23 |
| pt_max | 8·18 | 43·05 | 0·46 |
| pt_min | 8·18 | 43·05 | 0·46 |
| ptt_max | 8·43 | 57·03 | 0·46 |
| ptt_min | 8·43 | 57·03 | 0·46 |
| sodium_max | 0·25 | 3·44 | 0·23 |
| sodium_min | 0·25 | 3·44 | 0·23 |
| troponin_max | 66·31 | 92·16 | 0·46 |
| troponin_min | 66·31 | 92·16 | 0·46 |
| wbc_max | 0·89 | 4·39 | 0·23 |
| wbc_min | 0·89 | 4·39 | 0·23 |
| **Vital sign** | | | |
| cvp_max | 62·55 | 82·64 | 45·79 |
| cvp_mean | 62·55 | 82·64 | 45·79 |
| cvp_min | 62·55 | 82·64 | 45·79 |
| dbp_max | 0 | 0 | 0 |
| dbp_mean | 0 | 0 | 0 |
| dbp_min | 0 | 0 | 0 |
| fio2_max | 51·44 | 55·72 | 23·46 |
| fio2_min | 51·44 | 55·72 | 23·46 |
| gcs_max | 0 | 0 | 0 |
| gcs_mean | 0 | 0 | 0 |
| gcs_min | 0 | 0 | 1·14 |
| hr_max | 0·03 | 0·01 | 0 |
| hr_mean | 0·03 | 0·01 | 0 |
| hr_min | 0·03 | 0·01 | 0 |
| map_max | 0 | 0 | 0 |
| map_mean | 0 | 0 | 0 |
| map_min | 0 | 0 | 0 |
| rr_max | 0 | 0 | 0 |
| rr_mean | 0 | 0 | 0 |
| rr_min | 0 | 0 | 0 |
| sbp_max | 0 | 0 | 0 |
| sbp_mean | 0 | 0 | 0 |
| sbp_min | 0 | 0 | 0 |
| si_max | 0·13 | 0·33 | 0 |
| si_mean | 0·13 | 0·33 | 0 |
| si_min | 0·13 | 0·33 | 0 |
| spo2_max | 0·04 | 0 | 0 |
| spo2_mean | 0·04 | 0 | 0 |
| spo2_min | 0·04 | 0 | 0 |
| t_max | 0·09 | 0·02 | 0 |
| t_mean | 0·09 | 0·02 | 0 |
| t_min | 0·09 | 0·02 | 0 |
| **Clinical** | | | |
| dobutamine_flag | 0 | 0 | 0 |
| rate_dopamine_max | 0 | 0 | 0 |
| rate_epinephrine_max | 0 | 0 | 0 |
| rate_norepinephrine_max | 0 | 0 | 0 |
| vent_flag | 0 | 0 | 0 |
| crrt_flag | 0 | 0 | 0 |
| sirs_max | 0 | 0 | 0 |
| sirs_mean | 0 | 0 | 0 |
| sirs_min | 0 | 0 | 0 |
| sofa_neur_max | 0 | 0 | 0 |
| sofa_neur_mean | 0 | 0 | 0 |
| sofa_neur_min | 0 | 0 | 0 |
| sofa_resp_max | 0 | 0 | 0 |
| sofa_resp_mean | 0 | 0 | 0 |
| sofa_resp_min | 0 | 0 | 0 |
| uo_12hour | 3·83 | 31·1 | 4·1 |
| uo_24hour | 3·12 | 22·99 | 4·1 |
| uo_3hour | 30·81 | 57·48 | 6·15 |
| uo_6hour | 9·67 | 43·63 | 5·01 |

*alt alanine transaminase, ast aspartate transaminase, be the base excess, pafi PaO$_2$/FiO$_2$ ratio, pt the prothrombin time, ptt the partial thromboplastin time, si shock index, vent mechanical ventilation, crrt continuous renal replacement therapy, uo urine outout - sofa_neur explain*

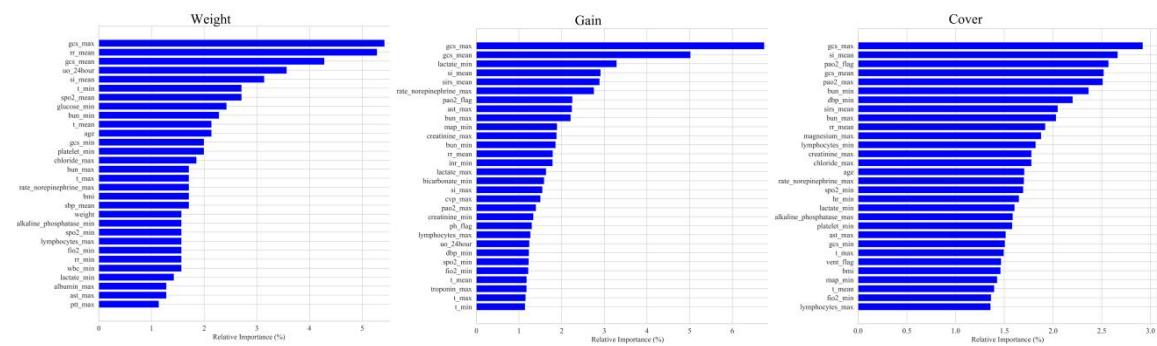

**eFigure 4: The importance ranking of top 30 risk factors utilizing the custom methods**

**eTable 5: The variables information of top 30 risk factors utilizing the custom methods**

| Weight | Gain | Cover |
| --- | --- | --- |
| gcs_max | gcs_max | gcs_max |
| rr_mean | gcs_mean | si_mean |
| gcs_mean | lactate_min | pao2_flag |
| uo_24hour | si_mean | gcs_mean |
| si_mean | sirs_mean | pao2_max |
| t_min | rate_norepinephrine_max | bun_min |
| spo2_mean | pao2_flag | dbp_min |
| glucose_min | ast_max | sirs_mean |
| bun_min | bun_max | bun_max |
| t_mean | map_min | rr_mean |
| age | creatinine_max | magnesium_max |
| gcs_min | bun_min | lymphocytes_min |
| platelet_min | rr_mean | creatinine_max |
| chloride_max | inr_min | chloride_max |
| bun_max | lactate_max | age |
| t_max | bicarbonate_min | rate_norepinephrine_max |
| rate_norepinephrine_max | si_max | spo2_min |
| bmi | cvp_max | hr_min |
| sbp_mean | pao2_max | lactate_min |
| weight | creatinine_min | alkaline_phosphatase_max |
| alkaline_phosphatase_min | ph_flag | platelet_min |
| spo2_min | lymphocytes_max | ast_max |
| lymphocytes_max | uo_24hour | gcs_min |
| fio2_min | dbp_min | t_max |
| rr_min | spo2_min | vent_flag |
| wbc_min | fio2_min | bmi |
| lactate_min | t_mean | map_min |
| albumin_max | troponin_max | t_mean |
| ast_max | t_max | fio2_min |
| ptt_max | t_min | lymphocytes_max |

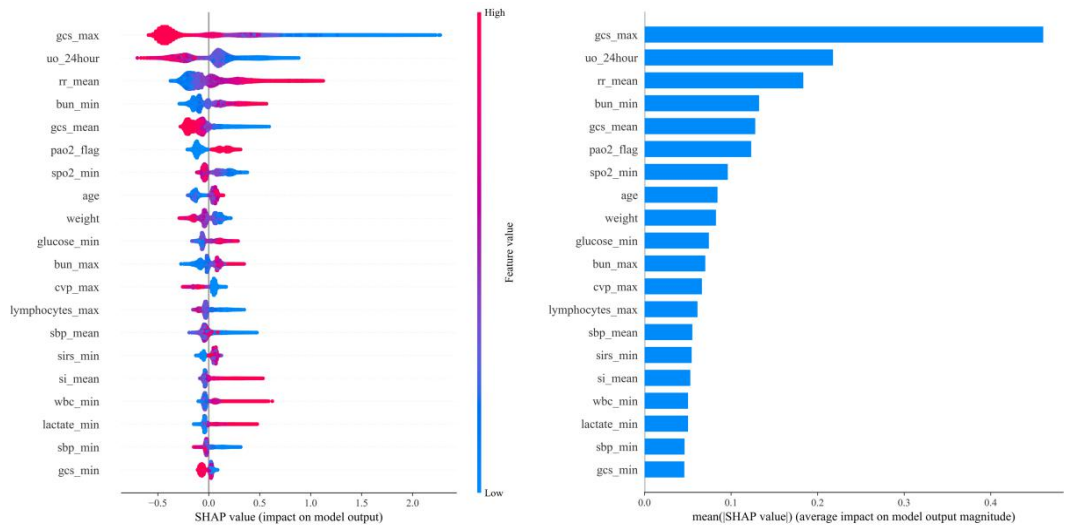

**eFigure 5: The importance ranking of the top 20 risk factors with stability and interpretation employing the MIMIC-III cohort**

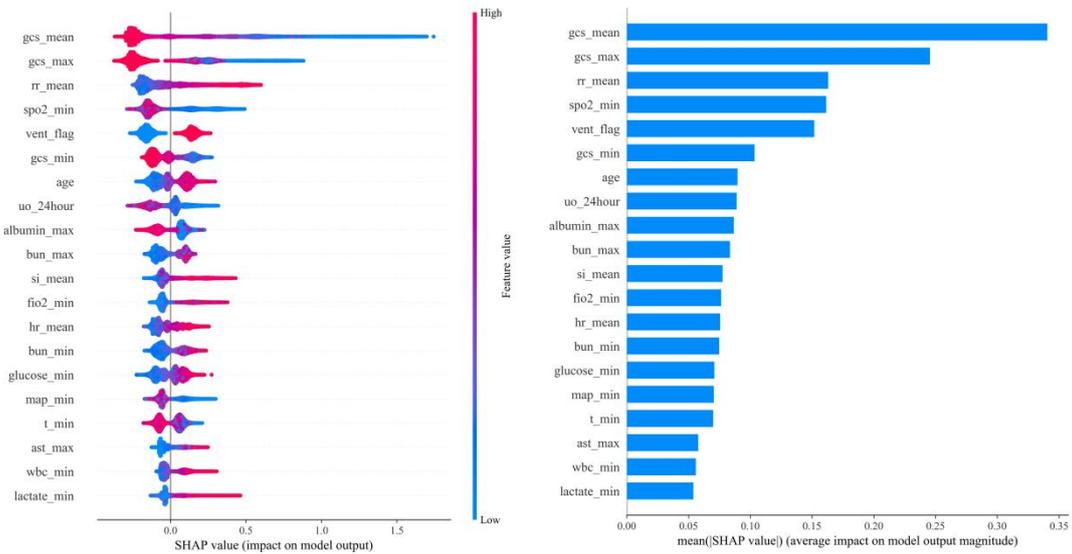

**eFigure 6: The importance ranking of the top 20 risk factors with stability and interpretation employing the eICU-CRD cohort**

**eTable 6: The variables information of top 20 risk factors with prediction models developing by different databases**

| MIMIC-III | eICU-CRD | MIMIC-III - eICU-CRD |
|---|---|---|
| gcs_max | gcs_mean | gcs_max |
| uo_24hour | gcs_max | gcs_mean |
| rr_mean | rr_mean | rr_mean |
| bun_min | spo2_min | bun_min |
| gcs_mean | vent_flag | age |
| pao2_flag | gcs_min | si_mean |
| spo2_min | age | sirs_mean |
| age | uo_24hour | gcs_min |

| weight | albumin_max | uo_24hour |
|---|---|---|
| glucose_min | bun_max | spo2_min |
| bun_max | si_mean | vent_flag |
| cvp_max | fio2_min | bun_max |
| lymphocytes_max | hr_mean | spo2_mean |
| sbp_mean | bun_min | bmi |
| sirs_min | glucose_min | glucose_min |
| si_mean | map_min | pao2_flag |
| wbc_min | t_min | lactate_min |
| lactate_min | ast_max | dbp_min |
| sbp_min | wbc_min | ast_min |
| gcs_min | lactate_min | pao2_max |

**eTable 7: The detailed information of the optimal model's performance evaluated in the MIMIC-III cohort**

|  | AUC (95% CI) | Sensitivity | Specificity |
|---|---|---|---|
| **Machine learning model** | | | |
| XGBoost (Our model) | 0·858 (0·841 - 0·875) | 0·834 | 0·705 |
| LR | 0·854 (0·838 - 0·871) | 0·845 | 0·712 |
| NN | 0·834 (0·817 - 0·853) | 0·784 | 0·735 |
| SVM | 0·824 (0·804 - 0·844) | 0·733 | 0·785 |
| RF | 0·788 (0·766 - 0·810) | 0·739 | 0·707 |
| NB | 0·741 (0·716 - 0·765) | 0·631 | 0·746 |
| **Clinical scores** | | | |
| OASIS | 0·752 (0·729 - 0·776) | 0·679 | 0·704 |
| APSIII | 0·73 (0·704 - 0·757) | 0·655 | 0·72 |
| MODS | 0·694 (0·669 - 0·719) | 0·849 | 0·417 |
| SAPS | 0·686 (0·659 - 0·713) | 0·494 | 0·771 |
| SOFA | 0·668 (0·639 - 0·697) | 0·429 | 0·838 |

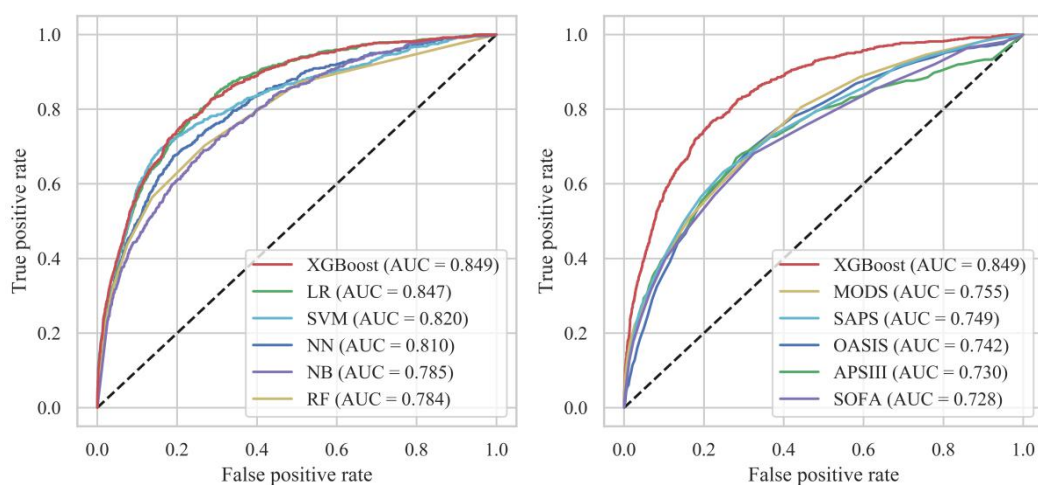

**eFigure 8: Our model's performance, evaluated in the eICU-CRD cohort, comparing with the**

eTable 8: The detailed information of the optimal model's performance evaluated in the eICU-CRD cohort

|  | AUC (95% CI) | Sensitivity | Specificity |
|---|---|---|---|
| **Machine learning model** | | | |
| XGBoost (Our model) | 0·849 (0·835 - 0·863) | 0·763 | 0·784 |
| LR | 0·847 (0·833 - 0·861) | 0·84 | 0·704 |
| NN | 0·81 (0·794 - 0·827) | 0·674 | 0·81 |
| SVM | 0·82 (0·803 - 0·838) | 0·696 | 0·838 |
| RF | 0·784 (0·766 - 0·803) | 0·703 | 0·731 |
| NB | 0·785 (0·768 - 0·802) | 0·69 | 0·734 |
| **Clinical scores** | | | |
| MODS | 0·755 (0·737 - 0·774) | 0·693 | 0·672 |
| SAPS | 0·749 (0·73 - 0·768) | 0·632 | 0·751 |
| OASIS | 0·742 (0·723 - 0·761) | 0·691 | 0·685 |
| APSIII | 0·730 (0·709 - 0·752) | 0·669 | 0·719 |
| SOFA | 0·728 (0·708 - 0·748) | 0·681 | 0·676 |

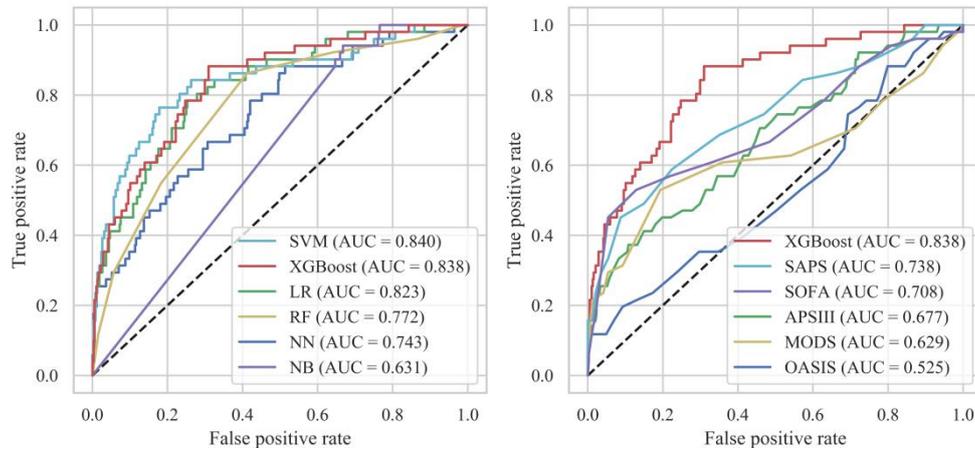

eFigure 9: Our model's performance, evaluated in the PLAGH-S cohort, comparing with the baseline models and clinical scores

eTable 9: The detailed information of the optimal model's performance evaluated in the PLAGH-S cohort

|  | AUC (95% CI) | Sensitivity | Specificity |
|---|---|---|---|
| **Machine learning model** | | | |
| SVM | 0·840 (0·774 - 0·906) | 0·765 | 0·822 |
| XGBoost (Our model) | 0·838 (0·780 - 0·895) | 0·882 | 0·691 |
| LR | 0·823 (0·762 - 0·883) | 0·784 | 0·75 |
| RF | 0·772 (0·705 - 0·839) | 0·863 | 0·588 |
| NN | 0·732 (0·670 - 0·815) | 0·667 | 0·727 |
| NB | 0·631 (0·588 - 0·674) | 0·941 | 0·332 |
| **Clinical scores** | | | |
| SAPS | 0·738 (0·658 - 0·818) | 0·588 | 0·776 |
| SOFA | 0·708 (0·619 - 0·796) | 0·529 | 0·871 |
| APSIII | 0·677 (0·595 - 0·759) | 0·373 | 0·892 |
| MODS | 0·629 (0·528 - 0·731) | 0·529 | 0·807 |
| OASIS | 0·525 (0·437 - 0·614) | 0·118 | 0·99 |

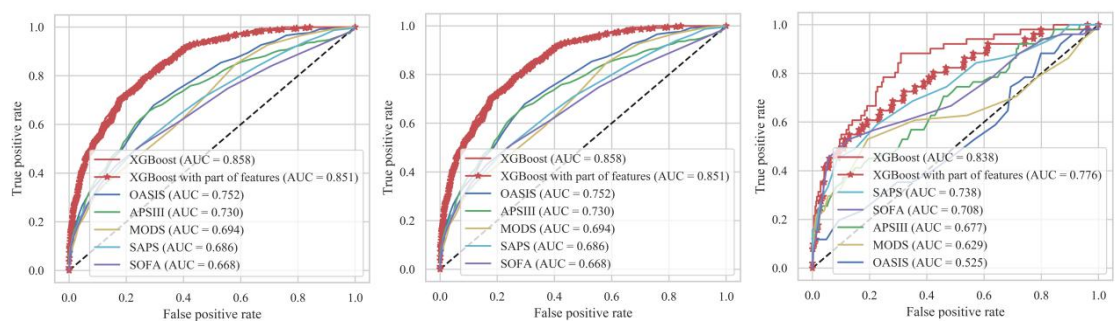

eFigure 10: The cross-validation of our model with the top 20 importance risk factors in the multicenter databases (MIMIC-III, eICU-CRD, PALGH-S)

eTable 10: The detailed information of our model's performance with cross-validation in the multicenter databases

|  | AUC | Sensitivity | Specificity |
|---|---|---|---|
| **MIMIC-III** | | | |
| Our model | 0·858 | 0·834 | 0·705 |
| Our model with part of features | 0·851 | 0·795 | 0·733 |
| OASIS | 0·752 | 0·679 | 0·704 |
| APSIII | 0·730 | 0·655 | 0·720 |
| MODS | 0·694 | 0·849 | 0·417 |
| SAPS | 0·686 | 0·494 | 0·771 |
| SOFA | 0·668 | 0·429 | 0·838 |
| **eICU-CRD** | | | |
| Our model | 0·849 | 0·763 | 0·784 |
| Our model with part of features | 0·839 | 0·712 | 0·814 |
| MODS | 0·755 | 0·693 | 0·672 |
| SAPS | 0·749 | 0·632 | 0·751 |
| OASIS | 0·742 | 0·691 | 0·685 |
| APSIII | 0·730 | 0·655 | 0·720 |
| SOFA | 0·728 | 0·681 | 0·676 |
| **PLAGH-S** | | | |
| Our model | 0·838 | 0·882 | 0·691 |
| Our model with part of features | 0·776 | 0·549 | 0·871 |
| SAPS | 0·738 | 0·588 | 0·776 |
| SOFA | 0·708 | 0·529 | 0·871 |
| APSIII | 0·677 | 0·373 | 0·892 |
| MODS | 0·629 | 0·529 | 0·807 |
| OASIS | 0·525 | 0·118 | 0·99 |

eTable 11: The additional information of cross-validation leveraging the total cohorts as the testing set

| Testing set \ Training set | | MIMIC-III | eICU-CRD |
|---|---|---|---|
| **MIMIC-III** | AUC | ·· | 0·823 |
| | Sensitivity | ·· | 0·774 |
| | Specific | ·· | 0·712 |
| | F1 | ·· | 0·624 |
| | Accuracy | ·· | 0·871 |
| **eICU-CRD** | AUC | 0·837 | ·· |
| | Sensitivity | 0·752 | ·· |
| | Specific | 0·772 | ·· |
| | F1 | 0·676 | ·· |
| | Accuracy | 0·891 | ·· |